\documentclass[aps,twocolumn,secnumarabic,nobalancelastpage,superscriptaddress,amsmath,amssymb,floatfix,nofootinbib]{revtex4}

\usepackage{graphics}      
\usepackage{graphicx}      
\usepackage{longtable}     
\usepackage{url}           
\usepackage{bm}            
\usepackage{epsfig}
\usepackage{amsmath}
\usepackage[usenames,dvipsnames]{xcolor} 
\begin{document}
\title{Thermal conductivity in one-dimensional nonlinear disordered lattices: Two kinds of scattering effects of hard-type and soft-type anharmonicities}
\author         {Jianjin Wang}
\email		   {phywjj@foxmail.com}
\affiliation    {Department of Physics, Jiangxi Science and Technology Normal University, Nanchang 330013, Jiangxi, China}
\author         {Chi Xiong}
\email		   {xiongchi@mju.edu.cn}
\affiliation    {MinJiang Collaborative Center for Theoretical Physics, College of Physics and Electronic Information Engineering, Minjiang University, Fuzhou 350108, China}
\author         {Daxing Xiong}
\email		   {xmuxdx@163.com}
\affiliation    {MinJiang Collaborative Center for Theoretical Physics, College of Physics and Electronic Information Engineering, Minjiang University, Fuzhou 350108, China}
\date{\today}

\begin{abstract}
The amorphous solids can be theoretically modeled by anharmonic disordered lattices. However, most of theoretical studies on thermal conductivity in anharmonic disordered lattices only focus on the potentials of hard-type (HT) anharmonicity. Here we study the thermal conductivity $\kappa$ of one-dimensional (1D) disordered lattices with both hard- and soft-type (ST) anharmonic on-site potentials. It is found, via both direct molecular dynamic simulations and theoretical method, that the anharmonicity dependence of $\kappa$ in the HT model is nonmonotonous, while in the ST model is monotonously increased. This provides a new way to enhance thermal conductivity in disordered systems. Furthermore, $\kappa$ of the HT model is consistent with the prediction of the quasi-harmonic Green-Kubo (QHGK) method in a wide range of anharmonicity, while for the ST model, the numerical results seem largely deviated from the theoretical predictions as the anharmonicity becomes soft. This new and peculiar feature of the ST model  may root in the fact that only delocalization effect exists, different from the competing roles that both delocalization and localization play in the counterpart HT model. 
\end{abstract}

\maketitle

\section{Introduction}

Due to the growing interest in predicting its value in various amorphous solids, the thermal conductivity, $\kappa$, of anharmonic disordered lattices has been extensively studied during the past few decades~\citep{PhysScr2018,Book2018,PMB1999,JAP2016,NaonoLett2016,ACS2011,AIPA2021,Slack1979, RPC1988,PM1972, JLTP1972,PRB1975,PRB1994, PRB1986,PRB1989,PRB1990,PRB1993,PRB1993-II,PRB2011,Science1991,PRL1994,JAP2009,NJP2016,NC2019,NP2019}. Nevertheless, even for the one-dimensional (1D) cases our understanding of the anharmonic disordered lattices with complicated potentials, in both theoretical and numerical viewpoints, is still far from complete~\citep{PhysScr2018,Book2018}.

For the ordered anharmonic crystal systems, the heat conductivity can be understood by the phonon gas model \citep{Peierls} where phonons are viewed as quasi particles performing random walks. Considering each phonon's lifetime or mean free path, Peierls~\citep{Peierls} first derived a formula, nowadays known as the Peierls formula, $\kappa=\frac{1}{3} C_V v^2 \tau$, where $C_V$, $v$, and $\tau$ are the specific heat capacity at constant volume, the group velocity, and the phonon lifetime, respectively. According to this theory, a correct description of temperature $T$ dependence of $\kappa$ at low temperatures can be obtained,  but only applicable to the 1D harmonic lattices with weak anharmonicity. It is thus worth mentioning that, for 1D lattice models with strong anharmonicity (with/without on-site potentials), both temperature-dependent behaviors at low and high temperatures have been predicted by the effective or self-consistent phonon theory (EPT or SCPT) ~\citep{PRB1991-JR,PRB1992-JR,PRE1993-Dauxois,PRE2008-He,PRE2007-Nianbei,PRE2014-Nianbei,PRE2021-He}. More challenging questions come from the study of the temperature/anharmonicity dependence of $\kappa$ for amorphous solids which are usually theoretically modeled by anharmonic disordered lattices, since the periodicity of this kind of systems is broken and the Anderson localization~\citep{PR1958} of phonons emerges. Consequently, some key physical concepts used in the previous studies, for example  the group velocity of phonons, are not applicable any more. Viewing this difficulty, over the past decades some theories for anharmonic disordered lattices, such as the minimum thermal conductivity theory~\citep{Slack1979, RPC1988}, the two-level states theory~\citep{PM1972, JLTP1972, PRB1975}, the energy hopping theory~\citep{PRB1994, PRB1986, PRB1989, PRB1990}, and the Allen-Feldman theory~\citep{PRB1993, PRB1993-II} have been developed and the limitations of these theories have been analyzed as well~\citep{PRB2011, Science1991, PRL1994, JAP2009, NJP2016}.

Recently,   a unified formula for predicting $\kappa$, based on the quasi-harmonic Green-Kubo (QHGK) method and applicable to both ordered and disordered anharmonic crystal systems, has been proposed and developed~\citep{NC2019, NP2019}. Comparing the predictions of this unified formula with the  corresponding results via molecular dynamics simulations, it was found that for a 1728-atom model of a-Si under $T < 600 $K,  the two methods provide consistent results while for $T > 600 $K, some deviations are observed~\citep{NC2019}. These deviations might  originate from that the unified formula is based on the harmonic approximation and the anharmonic effects are roughly  described by the decay rate of  non-interacting phonon modes. When the anharmonicity becomes strong, the nonlinearity  of the system not only brings  interactions among phonons but also alters the harmonic force-constant,  and hence, the frequencies and the configurations of phonon modes. This is also  the reason why in some previous studies~\citep{PRB1991-JR, PRB1992-JR, PRE1993-Dauxois, PRE2008-He,PRE2007-Nianbei,PRE2014-Nianbei,PRE2021-He}, to improve the accuracy of predictions, the SCPT are usually employed to calculate the effective harmonic force-constant induced by nonlinearity.

In this paper we perform a detailed study (via direct molecular dynamic simulations)  on the anharmonicity dependence of $\kappa$ in 1D disordered nonlinear lattices with two  distinctive one-site potentials, i.e., the hard-type (HT) and soft-type (ST) anharmoncities. We shall  first apply both SCPT and direct dynamic simulations to derive the effective harmonic force-constant,  which is then used in the QHGK method~\citep{NC2019} to predict thermal conductivity in both types of systems. Direct numerical results are obtained to examine the predictions. Our results demonstrate the  crucial difference between the disordered systems of HT and ST anharmonicities. Furthermore, by comparing the numerical results with the theoretical predictions,  consistence in the HT anharmonic disordered systems is found  while for the ST ones, large deviations are observed.  The results shows the limitation of the unified formula based on QHGK and provides insights for further improving the theoretical predictions and guiding the experimental studies of $\kappa$ in amorphous solids.

The rest of this article is organized as follows: In Sec. 2 we describe the two  types of disordered model systems with the HT and ST on-site potentials. Section 3 presents the main results of computing heat conductivity.  We first show the anharmonicity dependence of $\kappa$ from the numerical simulations and the theoretical QHGK method, respectively~\citep{NC2019},  then study the effective force-constant via both SCPT and numerical examinations, the decay rate and the localization properties of the Anderson normal modes. Key differences of the two types of systems  (HT and ST) are revealed  and for the latter case, the discrepancy between the unified formula in QHGK and the direct molecular simulations are discussed. Finally,  our results are summarized and conclusions are drawn in Sec. 4.
\section{Model}
To model a 1D amorphous solid, we start with a 1D mass disordered harmonic system with an anharmonic on-site potential. Its dimensionless Hamiltonian reads~\citep{PRE1996}
\begin{equation}
\label{HH}
H=\sum_{i}^N\frac{p_i^2}{2m_i}+\frac{1}{2}(x_{i+1}-x_i)^2+\frac{1}{2}\frac{(x_i^2+ \xi x_i^4)}{1+x_i^2},
\end{equation}
where $x_i$ is the displacement of the $i$th particle from its equilibrium position and $p_i$ is its momentum. $m_i$ is the mass of each particle which is a random quantity disordered systems and set uniformly distributed in an interval of $[0.8, 1.2]$. $N$ (always set to be $4096$ in the following) is the total number of particles in the disordered lattice. In the summation of the Hamiltonian, the first, second and third terms correspond to the kinetic energy $E_i$, the interparticle potential $V$ and the on-site potential $U_i$, respectively. In particular, in the on-site potential, $\xi$ is a controlled parameter to capture the features of on-site potential depending on whether it is an HT or ST type. From the series expansion of $U_i$ at $x_i=0$
\begin{equation}
U_i \propto 0.5 x_i^2+(0.5 \xi-0.5) x_i^4+(0.5-0.5 \xi) x_i^6 +O(x_i^8),
\end{equation}
one can find that, in the case of $\xi=1$, the system reduces to a pinned-harmonic system where the phonon-phonon interactions are absent. That means that all the phonon modes are Anderson localized ones~\citep{PRE2016}. However, for $\xi >1$ ($\xi <1$), the HT (ST) feature emerges due to the contributions of the higher order terms. Therefore, in this model one can study two types of disordered systems with both HT and ST anharmonicities only by adjusting the value of $\xi$.
\section{Results and discussions}
As usual we let the system evolve dynamically according to the canonical equations of  the aforementioned Hamiltonian with the help of numerical integral algorithm. In our molecular dynamic simulations, the velocity-Verlet scheme~\citep{Aullen1987} with an integral step $0.01$ is always applied. This ensures that the error of total energy density should be at least in an order of $10^{-5}$. Unless otherwise specified, we set the  equilibrium temperature of the system to be $T=0.5$ and impose periodic boundary conditions.
\subsection{Thermal conductivity}
With the above model and simulation setups, we now are able to study the $\xi$-dependence of thermal conductivity $\kappa$. Numerically, there are two main approaches to achieve this. The first one is based on the Green-Kubo formula~\citep{GK1, GK2, Kubo1991}
\begin{equation}
\kappa=\lim\limits_{\tau \rightarrow \infty }\lim\limits_{N\rightarrow
\infty} \frac{1}{k_{B}T^{2}N}\int_{0}^{\zeta}C(t)dt,
\label{GK}
\end{equation}
where $k_B$ is the Boltzmann constant (set to be $1$), $C_{JJ}(t)=\langle J(0)J(t)\rangle$ is the time autocorrelation of heat currents, where $J=\sum_i\frac{1}{2}(\dot{x}_{i+1}+\dot{x_i})\frac{\partial H(x_{i+1}-x_i)}{\partial x_i}=\sum_i-\frac{1}{2}(\dot{x}_{i+1}+\dot{x}_i)(x_{i+1}-x_i)$  is the total heat current [$\dot{x}_i=\frac{dx_i(t)}{dt}$]~\citep{PR2003, AP2008} and $\langle\cdot\rangle$ denotes the ensemble average. To obtain $C_{JJ}(t)$, the system is first thermalized to a given temperature with the Langevin heat baths, which are then removed after the equilibrium state is eventually reached. Finally, $\kappa$ can be measured through the Green-Kubo integration from Eq.~(\ref{GK}). For each $C_{JJ}(t)$, we take average over $24$ different initial conditions.
\begin{figure*}[]
\centering
\includegraphics[scale=0.6]{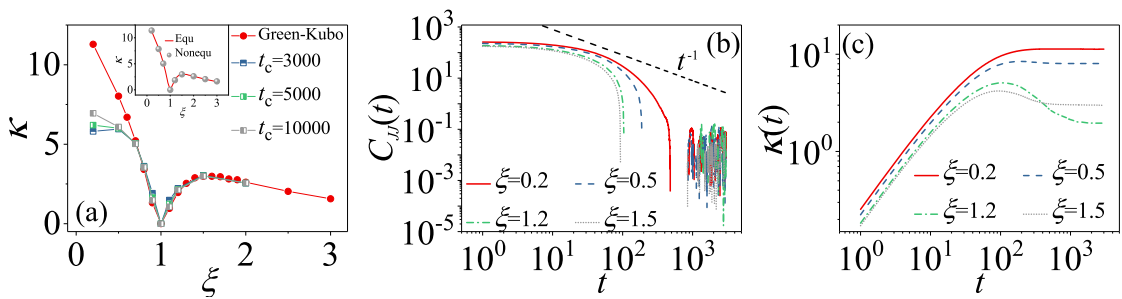}
\caption{(a) Results of the $\xi$-dependence of $\kappa$, where the circles (with the line as a guide, the same below) are the results ofcalculating $\kappa$ from the Green-Kubo formula, which showquantitatively the same results with those derived by the nonequilibrium method (see the inset); the half-opened squares denote the results of $\kappa$ predicted by the unified formula~(\ref{NC}). (b) $C_{JJ}(t)$ for different $\xi$ in the equilibrium approach. (c) The corresponding integration of $C_{JJ}(t)$ with $t$.}\label{fig1}
\end{figure*}

The second method is from the direct molecular dynamic simulations of nonequilibrium states. In this method,  two ends of the system are first connected to two Noose-Hoover heat baths with temperatures $T_+=0.6$ and $T_{-}=0.4$ (the average temperature of the system is then $T=0.5$), respectively. These thermal baths will drive the system to a nonequilibrium steady state, and also produce a well-behaved temperature gradient that results in heat current flowing across the system. Finally, the thermal conductivity $\kappa$ can be calculated through the Fourier law $\kappa=jN/(T_+-T_- )$ with $j$ being the heat current density flowing through each particle.

Figure~\ref{fig1}(a) shows our main results of $\xi$-dependence of $\kappa$. As can be seen in the inset, both the equilibrium and nonequilibrium approaches give quantitatively the same results. First, the harmonic disordered model ($\xi=1$) has a minimal $\kappa=0$ because in this case all the phonon modes become Anderson localized as expected. Second, in the HT type model a usual nonmonotonous $\xi$-dependence of $\kappa$ can be identified~\citep{Ourwork2020}, i.e., as $\xi$ increases from $\xi=1$ (the on-site potential becomes hard), $\kappa$ first increases, then reaches its maximum at certain $\xi$ value, and finally decreases. This nonmonotonous behavior has been attributed to the delocalization and re-localization of the Anderson modes induced by the combined effects of disorder and nonlinearity~\citep{Ourwork2020}. Bearing this in mind, let us now turn to the results of the ST type model. Interestingly, as $\xi$ decreases from $\xi=1$, i.e., the on-site potential becomes soft, the nonmonotonous behavior no longer exists. $\kappa$ monotonously increases with the decrease of $\xi$, implying that only delocalization processes appear in the ST systems, which is quite different from the counterpart HT systems.

Some results from the equilibrium approach, e.g. the heat current auto-correlations $C_{JJ}(t)$ of $\xi=0.2, 0.5, 1.2$, and $1.5$ are shown in Fig.~\ref{fig1}(b). All of the correlations decay faster than $t^{-1}$, indicating the possible convergence to derive $\kappa$ by using the Green-Kubo formula. This is also consistent with the known results of normal heat conduction in momentum nonconserving systems obeying the Fourier law~\citep{PR2003, AP2008, Zhao1998}. The corresponding integration of Eq.~(\ref{GK}) for each $\xi$ is presented in Fig.~\ref{fig1}(c), which further confirms our conjecture of the convergence.

\subsection{Predictions by the unified formula}
We next present the predictions by the unified formula~\citep{NC2019, NP2019} to see if they can be validated by our 1D disordered models with both HT and ST on-site potentials, and especially how the ST anharmonicity plays a role in the unified formula. For facilitating the comparison with the simulation results, here we only focus on the unified formula based on QHGK method proposed in~\citep{NC2019}. When applying to the 1D model the formula reads:
\begin{align}
\label{NC}
\kappa&=\frac{k_B}{N}\sum_{nm}v_{nm}^2\tau_{nm}, \nonumber \\
v_{nm}&=\frac{1}{2\sqrt{\omega_n\omega_m}}\sum_{ij}\frac{R_i^o-R_j^o}{\sqrt{m_im_j}} \Phi_{ij}e_n^i e_m^j, \\
\tau_{nm}&=\frac{\gamma_n+\gamma_m}{(\gamma_n+\gamma_m)^2+(\omega_n-\omega_m)^2}; \nonumber
\end{align}
where $v_{nm}$ and $\tau_{nm}$ are the generalized group velocity and the lifetime of Anderson modes, respectively; $\omega_n$, $e_n$, and $\gamma_n$ is the frequency, eigenvector, and decay rate of the $n$th normal mode, respectively; $R_i^o$ is the equilibrium position of the $i$th particle, and $\Phi_{ij}=(\frac{\partial^2 H}{\partial x_i\partial x_j})_0$ is the element of the force-constant matrix (here the subscript $0$ indicates the equilibrium position of the reference particle).
\begin{figure*}[]
\centering
\includegraphics[scale=0.6]{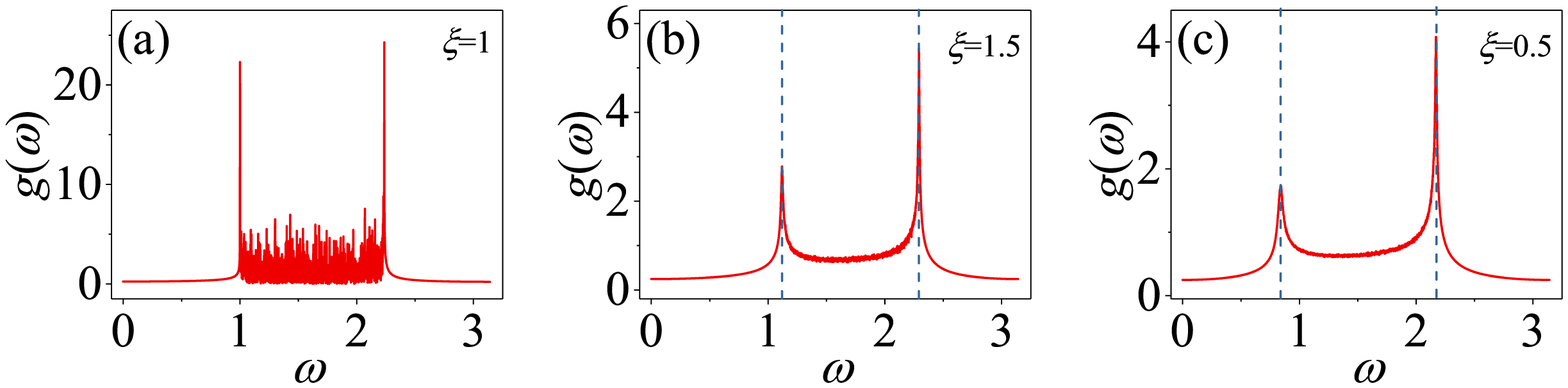}
\caption{DOS for $\xi=1$ (a), $1.5$ (b), and $0.5$ (c) of the corresponding homogeneous systems.}
\label{fig2}
\end{figure*}

To implement this formula, one may simply set $x_i=A_i \exp(- \rm i \mathrm{\omega_n} \mathrm{t})$, which enables us to linearize the motion equation of particles such that the dynamic matrix can be diagonalized. Let $e_n^i$ and $\omega_n$ be the $n$th normalized eigenvector and eigenvalue of the dynamic matrix, and the projection of the motion onto the $n$th Anderson mode be
$Q_n(t)=\sum_i^N \sqrt{m_i} x_i(t) e_n^i$ and $P_n(t)=\sum_i^N \sqrt{m_i} \dot{x}_i(t) e_n^i$ with $n=1, \ldots, N$ and $P_n$ and $Q_n$ being the canonical momentum and canonical coordinate, respectively. Here we only present the predictions of formula~(\ref{NC}) in Fig.~\ref{fig1}(a), leaving the details of obtaining  $\Phi_{ij}$, $\gamma_n$, etc. for calculating $\kappa$ in the following sections.

As shown in Fig.~\ref{fig1}(a), the unified formula~(\ref{NC}) gives an accurate prediction of $\kappa$ for the HT disordered model even when the system is strongly anharmonic, for example $\xi=2$. However, for the ST disordered model, this formula only holds for the range $0.7\leqslant \xi<1.0$. The predicted values of $\kappa$ seem to be lower than those from direct dynamic simulations when $\xi < 0.7$, even if we have further increased the integration time to obtain the decay rate of the Anderson normal modes [see Fig.~\ref{fig1}(a) and the detailed calculations below]. These results further demonstrate that the ST anharmonic disordered systems is beyond the scope of the unified formula.

\subsection{Effective harmonic force-constant}
The matrix $\Phi_{ij}$ in the QHGK method~\citep{NC2019} is composed of force-constants. However, as mentioned in introduction, one cannot use the origin harmonic force-constant to construct the dynamic matrix.  Instead an effective force-constant $k_{eff}$, induced by both HT and ST anharmonicities, should be first given when applying the QHGK method. Here we present the analytical results of SCPT for deriving $k_{eff}$, and then provide the numerical measure from direct dynamic simulations.

The basic idea of SCPT is to use a harmonic Hamiltonian $H_0$ to approximate a general Hamiltonian $H$. Then,  by minimizing the free energy, one can get a self-consistent equation for $k_{eff}$~\citep{PRB1991-JR, PRB1992-JR, PRE1993-Dauxois, PRE2008-He,PRE2007-Nianbei,PRE2014-Nianbei,PRE2021-He}. Considering a  general Hamiltonian $H$, SCPT tells us that one can use an equivalent harmonic Hamiltonian $H_0$ so that the free energy is minimized, i.e.
\begin{equation}
F\leqslant F_0+\langle H-H_0\rangle_0
\label{free}
\end{equation}
where $F$, $F_0$, and $\langle H-H_0\rangle _0$  are the free energies of the original system, the effective harmonic system, and the average of their Hamiltonian difference, respectively.  Note that the average is taken over the equivalent harmonic system and the minimization is taken on $\langle H-H_0\rangle _0$. Further, considering the statistical equivalence of particles in lattice systems, it is actually feasible to minimize $\langle H-H_0\rangle _0$ for a single particle.

For our focused Hamiltonian~(\ref{HH}), the anharmonicity only appears in the on-site potential, and $H_0=\sum_i\frac{p_i^2}{2m_i }+\frac{1}{2}(x_{i+1}-x_i )^2+\frac{k_{eff}}{2} x_i^2$. To calculate $\langle H-H_0\rangle _0$, we first write the on-site potential in Hamiltonian~(\ref{HH}) in an inverse form
\begin{equation}
	\begin{split}
		 \langle H-H_0\rangle _0 &=\left \langle\frac{1}{2}\frac{1}{\frac{1}{x^2}+\frac{1-\xi}{1+\xi x^2}}-\frac{1}{2} k_{eff} x^2 \right \rangle_0\\
		&=-\frac{1}{2} k_{eff} \langle x^2\rangle_0+\frac{1}{2}\frac{1}{\frac{1}{\langle x^2\rangle_0}+\frac{1-\xi}{1+\xi \langle x^2\rangle_0}},
	\end{split}
\end{equation}
where $\langle x^2\rangle_0=\frac{\int x^2e^{-\frac{k_{eff}}{2 k_B T}  x^2}dx}{\int e^{-\frac{k_{eff}}{2 k_B T} x^2}dx}=\frac{T}{k_{eff}}$. Now let $y=\langle x^2\rangle_0$, $\langle H-H_0\rangle _0$ can be rewritten as $f=-\frac{1}{2}\xi y+\frac{1}{2}\frac{1}{\frac{1}{y}+\frac{1-\xi}{1+\xi y}}$. To satisfy the Eq.~(\ref{free}),  one requires that $\frac{\partial f}{\partial y}=0$, from which a self-consistent equation $k_{eff}^3+(2T-1) k_{eff}^2+(T^2-2\xi T)k_{eff}-\xi T^2=0$ can be obtained. Finally  substituting the given temperature $T=0.5$ into this self-consistent equation, one obtains an explicit self-consistent relation between $k_{eff}$ and $\xi$:
\begin{equation}
k_{eff}^3+(\frac{1}{4}-\xi)k_{eff}-\frac{1}{4}\xi=0.
\label{SCPT}
\end{equation}

Eq.~(\ref{SCPT}) provides an analytical relation to derive $k_{eff}$. Here  it is also desirable to find $k_{eff}$ directly from simulations. For this purpose, we now consider the density of states (DOS) $g(\omega)$ of the system. As it is well known that in a pinned harmonic homogeneous system [with Hamiltonian $H=\sum_{i}^N\frac{p_i^2}{2}+\frac{1}{2}(x_{i+1}-x_i)^2+k_{eff} \frac{1}{2} x_i^2$], the minimal and maximal frequencies are $\omega_{min}=\sqrt{k_{eff}}$ and $\omega_{max}=\sqrt{4+k_{eff} }$, respectively. This helps us obtain $k_{eff}$ numerically. In practice, DOS can be obtained via Fourier transformation of the particle velocity auto-correlation. Some typical results of DOS for different $\xi$ are shown in Fig.~\ref{fig2}. For a harmonic system [$\xi=1$, see Fig.~\ref{fig2}(a)], there are  actually two peaks  located at $\omega=1$ and $\omega \simeq 2.236$, respectively, indicating $k_{eff}=1$. For the HT anharmonic systems [for example $\xi=1.5$, see Fig.~\ref{fig2}(b)], the frequency shift of the phonon modes happens, while in the case of ST anharmonicsystems [e.g., $\xi=0.5$, see Fig.~\ref{fig2}(c)], the softening of the normal modes (the decrease of $\omega_n$) takes place. Both evidences enable us to measure $k_{eff}$  caused by anharmonic on-site potential for the corresponding disordered systems.

\begin{figure}[]
\centering
\includegraphics[scale=0.25]{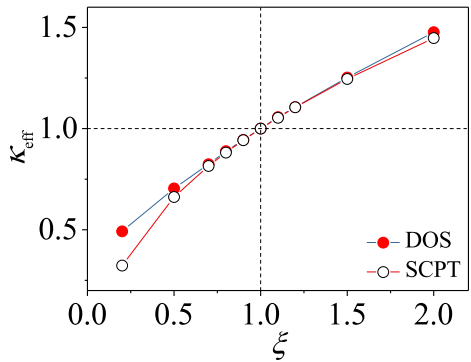}
\caption{$k_{eff}$ versus $\xi$ from SCPT (open) and from our direct simulation measurements based on the DOS (solid).}
\label{fig3}
\end{figure}

In Fig.~\ref{fig3} we compare the results of $k_{eff}$ for different $\xi$ both from SCPT and from our direct simulation measurements based on DOS. As can be seen, in a fairly wide range of $\xi$ around $\xi=1$, both  approaches give consistent results. However, as the on-site potential becomes soft, some  discrepancy can be identified.  This discrepancy becomes more significant when we look at the ST model (e.g. from $\xi=0.5$ and below, Fig.~\ref{fig3}).  Therefore for a more complicated ST anharmonic disordered model system, one should be more cautious when applying the unified formula to predict $\kappa$. Viewing all of these, in the following we shall include $k_{eff}$ from our direct dynamic simulation measurements  in the practical calculations in Eq.~(\ref{NC}).
\subsection{Decay rate of the Anderson modes}
\begin{figure}[]
\centering
\includegraphics[scale=0.45]{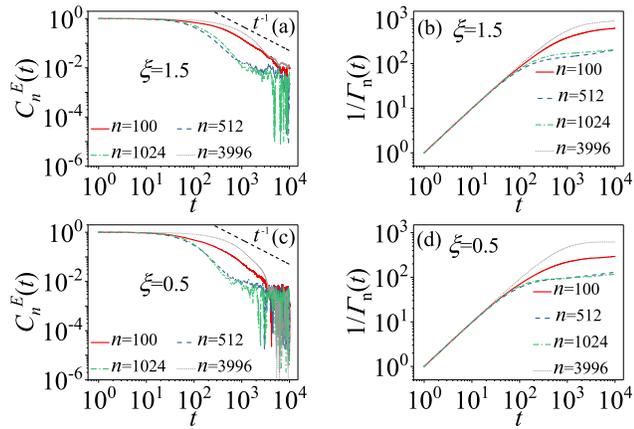}
\caption{(a) and (c): $C_n^E(t)$ versus $t$ for the $n$th Anderson modes of $n=100$, $512$, $1024$, and $3996$, respectively, for $\xi=1.5$ and $\xi=0.5$. (b) and (d): The corresponding inverse of $\Gamma_n$ versus time $t$.}
\label{fig4}
\end{figure}

With $k_{eff}$, the eigenvalue $\omega_n$ and eigenvector $e_n$ of an Anderson mode can be immediately extracted from  diagonalizing the dynamic matrix. For applying the QHGK formula~\citep{NC2019}, we still need to know the decay rate $\gamma_n$.  In dynamic simulations, $\gamma_n$ can be measured by the  time evolution of the equilibrium energy-energy autocorrelation of a single Anderson mode, defined by
\begin{equation}
C_n^E (t)=\frac{\langle\Delta E_n (0)\Delta E_n (t)\rangle}{\langle\Delta E_n (0)\Delta E_n(0)\rangle},
\label{cn}
\end{equation}
where $\Delta E_n (t)=E_n (t)-\langle E_n\rangle$ with $E_n=\frac{1}{2} (P_n^2+\omega_n^2 Q_n^2 )$ being the instantaneous energy of the Anderson mode $n$. Therefore, under the single-mode relaxation approximation, the $n$th mode energy decay rate $\Gamma_n$~\citep{PRB1986-Ladd}  is determined by
\begin{equation}
\frac{1}{\Gamma_n}=\lim\limits_{t_c\rightarrow\infty}\int_0^{t_c}C_n^E(t)dt.
\label{Lifetime}
\end{equation}
 Note that mathematically $\gamma_n=\Gamma_n/2$ since in principle the $n$th mode decay rate $\gamma_n$ should be measured by a similar definition $\frac{1}{\gamma_n}=\lim\limits_{t_c\rightarrow\infty}\int_0^{t_c}C_n^Q(t)dt$ with $C_n^Q (t)=\frac{\langle\Delta Q_n (0)\Delta Q_n (t)\rangle}{\langle\Delta Q_n (0)\Delta Q_n (0)\rangle}$.  Measuring $C_n^E(t)$, however,  is just for facilitating the numerical simulations.

The inverse of $\Gamma_n$ is called the lifetime of the $n$th Anderson mode. To obtain a finite $\Gamma_n$ from the integration in Eq.~(\ref{Lifetime}), $C_n^E(t)$ should be a fast decay function. If the decay of $C_n^E(t)$ were not fast enough, the integration would have become divergent, indicating an infinite lifetime of the normal mode. This means that there is always a part of energy fluctuations locked in the normal localized mode. Given that, in practice it is better to truncate the integration in Eq.~(\ref{Lifetime}) at some time cutoff $t_c$ (see also Fig.~\ref{fig1}) and analyze how the integration changes.
\begin{figure}[]
\centering
\includegraphics[scale=0.45]{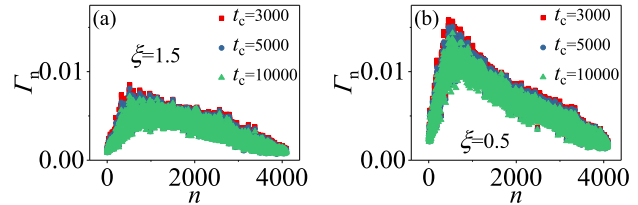}
\caption{ $\Gamma_n$ versus mode's index $n$ at a truncated time $t_c=3000$, $5000$, and $10000$ for (a) $\xi=1.5$ and (b) $\xi=0.5$.}
\label{fig5}
\end{figure}

Figures~\ref{fig4}(a) and (c) depict the decay of $C_n^E(t)$ for four typical Anderson modes. Fortunately all these $C_n^E(t)$ decay faster than $t^{-1}$, indicating a finite lifetime of these modes. These have been numerically verified in Figs.~\ref{fig4}(b) and (d). Some observed details are noteworthy, e.g. $C_n^E(t)$ with $n$ being some number in the middle of mode range (e.g., $n=512$ and $1024$) decays faster than those with $n$ at boundaries (e.g., $n=100$ and $3996$ in our simulations). This indicates that the $512$th and $1024$th Anderson modes have shorter lifetimes than those of the $100$th and $3996$th modes, which is consistent with the previous results observed only in the HT models~\citep{Ourwork2020}.

To further study the properties of  the ST anharmonic disordered systems, we provide  $\Gamma_n$ in Fig.~\ref{fig5} for three truncated times $t_c$. On the one hand, for all modes, $\Gamma_n$ nearly do not change with $t_c$, indicating possible convergence of the integration~(\ref{Lifetime}); On the other hand, from the results of Figs.~\ref{fig5}(a) and (b), the lifetimes of the Anderson modes for the ST anharmonic disordered systems seem shorter than those of the HT ones, implying a stronger delocalization induced by the ST anharmonicities.
\subsection{Discussions: Participation number and localization length}
\begin{figure}[]
\centering
\includegraphics[scale=0.45]{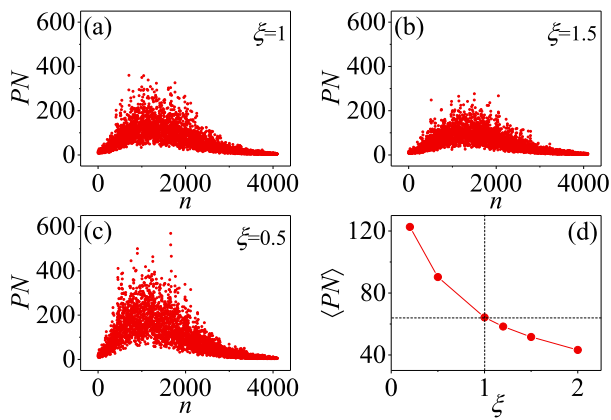}
\caption{Participation numbers $PN$ versus mode's index $n$ for (a) $\xi=1$, (b) $\xi=1.5$, and (c) $\xi=0.5$. (d) The averaged participation number $\langle PN \rangle$ over $n$ versus $\xi$.}
\label{fig6}
\end{figure}

From the above direct dynamic simulations via both the equilibrium and nonequilibrium approaches,  plus the theoretical QHGK unified formula~\citep{NC2019}, we know that the ST anharmonic disordered systems have larger thermal conductivity $\kappa$ than the counterpart HT systems. In addition, the current QHGK method~\citep{NC2019}, even including the effective force-constant that obtained both from SCPT and direct simulations, is still not able to give a thorough prediction for the ST anharmonic disordered systems. The distinctions from the counterpart HT systems and the invalidity of the predictions seem to be relevant to the same invalidity of SCPT and the shorter lifetime of the Anderson modes in the ST anharmonic disordered systems. To further investigate why the aforementioned methods fail to calculate $\kappa$ of the ST systems, we discuss the localization properties of the normal Anderson modes, characterized by their localization lengths.

The localization properties of phonon modes can be studied by the participation number $PN$ which is defined by $PN=\frac{(\sum A_i^2)^2}{\sum A_i^4}$~\citep{PN1970}, where $A_i$ is the amplitude of the displacement $x_i=A_i \exp(- \rm i \mathrm{\omega_n} \mathrm{t})$ that we have already substituted into the equations of motion to diagonalize the dynamic matrix. For a normal mode ($\omega_n$) in which $A_i$ is constant, $PN=N$, which implies a uniform participation of motions for all particles in that mode. This is a sign of delocalization. For a mode whose motions are carried by a single particle, $PN=1$ which indicates the highest degree of localization. Equivalently, $PN$ can be considered as the localization length of the modes and is inversely proportional to the degree of localization since a mode with a smaller localization length corresponds to a higher degree of localization.

Figure~\ref{fig6} depicts $PN$ versus the mode index $n$ for $\xi=1$, $\xi=1.5$, and $\xi=0.5$, respectively. Note that in practice we only need to calculate $\frac{1}{\sum A_i^4}$ to derive $PN$ due to the normalization of eigenstates $\sum A_i^2=1$. As can be seen in Figs.~\ref{fig6}(a)-(c), the curves of $PN$ versus $n$ are bell-shaped, i.e., the Anderson mode with a middle $n$ value has a relatively larger $PN$ than those with $n$ at boundaries. This implies that the middle $n$th mode of the corresponding homogeneous systems has a larger group velocity than the boundary modes~\citep{PSheng}. Furthermore, a detailed comparison of Figs.~\ref{fig6}(a)-(c) shows that the HT potential enhances the Anderson localization while the counterpart ST potential mainly weakens it. From Figs.~\ref{fig6}(b) and (c), $PN$ of the ST model is manifestly larger than that of the HT model. This is also supported by the results of the averaged $PN$ ($\langle PN\rangle$) with respect to $\xi$ in Fig.~\ref{fig6}(d). This is a crucial distinction between the HT and ST model systems. Such observations are also consistent with the decay rates of localized normal modes as shown in Fig.~\ref{fig5}.

Now let us understand more about the results of thermal conductivity shown in Fig.~\ref{fig1}(a). For the purely harmonic disordered systems, all the Anderson modes are fully localized and non-interacting, hence $\kappa=0$. For the HT anharmonic disordered systems, on the one hand the HT anharmonic on-site potential brings mode-mode interactions into the system; on the other hand it eventually enhances the localization. Therefore, these two mechanisms compete and finally lead to the nonmonotonous finite $\xi$-dependent $\kappa$. Here $\kappa$ approaches a finite value and can be accurately predicted by the unified formula~(\ref{NC}).  The scenario, however, changes dramatically for the ST anharmonic on-site potential, whose main effect on disordered systems is to delocalize the Anderson modes, similar to the effect of the mode interactions induced by nonlinearity. Combining both effects it results in a stronger monotonous increase of $\kappa$ with respect to $\xi$.  As we at present do not know whether there is a upper limit of $\kappa$, and $\kappa$ becomes larger when the anharmonicity keeps increasing, it is hard to predict $\kappa$ by any existing theories when $\xi$ exceeds a certain threshold value.
\section{Conclusion}
To summarize, by employing a 1D disordered lattice systems with both HT and ST anharmonic on-site potentials controlled by an adjustable parameter $\xi$, we have shown that different types of anharmonicities can result in quiet distinct nonlinearity dependence of thermal conductivity $\kappa$. The usual harmonic disordered system ($\xi=1$) has a zero $\kappa$  due to the absence of anharmonicity, thus only fully localized Anderson modes survive in the system. Taking this as a reference point, as the on-site potential becomes hard ($\xi >1$)  we observe a nonmonotonous $\xi$-dependent behavior of $\kappa$. However,  for ($\xi <1$) of the ST anharmonic case, we only see a monotonous increase of $\kappa$. This is an important finding in our present work, and as far as we know, for the first time demonstrated the different scattering roles that HT and ST anharmonicities play in disordered systems.

We have also carefully examined the validity of the recently proposed QHGK unified formula~\citep{NC2019} in our model. In comparison with the direct dynamic simulations, it is found that $\kappa$ of the HT model is consistent with the prediction of the QHGK method for a wide range of $\xi$, while in the ST model, the numerical results of $\kappa$ seem largely deviated from the theoretical predictions as the anharmonicity becomes soft. This is even the case when we include the effective force-constant induced by nonlinearity into the unified formula. It suggests that more theoretical efforts are needed to understand the thermal conductivity in amorphous solids with complicated interactions.

To explore the peculiarity on localization of the ST anharmonic disordered systems and answer why the ST potentials can enhance thermal conductivity of disordered systems, we have further studied  the localization properties of the Anderson normal modes, and revealed the fact that while delocalization and localization induced by the  HT anharmonicity compete with each other, the ST anharmonicity can only delocalize the Anderson modes. This new finding of purely delocalization mechanism naturally shed light on paving a new way to enhance the thermal conductivity in amorphous solids.
\begin{acknowledgments}
J.W. is supported by NNSF (Grant No. 12105122) of China; C.X. is supported by NSF (Grant No. 2022J011130) of Fujian Province of China; D.X. is supported by NNSF (Grant No. 12275116) of China and NSF (Grant No. 2021J02051) of Fujian Province of China.
\end{acknowledgments}

\end{document}